\documentclass[a4paper,fleqn,usenatbib,useAMS]{mnras}
\usepackage{newtxtext,newtxmath}
\usepackage[T1]{fontenc}
\usepackage{ae,aecompl}
\usepackage{graphicx}    
\usepackage{amsmath}     
\usepackage{multicol}    
\usepackage{bm}          
\usepackage{pdflscape}   
\usepackage{xspace}
\newcommand{\msun}{\,$M_{\sun}$}

\newcommand{\ergs}{\,erg\,s$^{-1}$}
\newcommand{\kms}{\,km\,s$^{-1}$}

\newcommand{\cmq}{\,cm$^{-3}$}
\newcommand{\gcmq}{\,g\,cm$^{-3}$}

\newcommand{\ha}{\,H$\alpha$}
\title[SN 2008iy circumstellar interaction] 
  {SN 2008iy circumstellar interaction: Bright and lesser light effects
   }
\author[N. N. Chugai]{
N. N. Chugai\thanks{E-mail: nchugai@inasan.ru}
\\
$^{1}$Institute of Astronomy, Russian Academy of Sciences, Pyatnitskaya
      St. 48, 119017 Moscow, Russia\\
}

\date{Accepted XXX. Received YYY; in original form ZZZ}

\pubyear{2021}

\begin{document}
\label{firstpage}
\pagerange{\pageref{firstpage}--\pageref{lastpage}}
\maketitle
%
\begin{abstract}
Optical photometry and spectra of the luminous type IIn supernova SN~2008iy are analysed in 
   detail with implications for cosmic ray acceleration and the radio emission. 
The light curve and expansion velocities indicate ejecta with the kinetic energy of $3\times10^{51}$\,erg    
  to collide with the $\sim10$\msun\ circumstellar envelope. 
The luminous \ha\ is explained as originated primarily from circumstellar clouds 
  interacting with the forward shock. 
For the first time the fluorescent O\,I 8446\AA\ emission is used to    
   demonstrate that the cloud fragmentation cascade spans a scale range $> 2.3$\,dex. 
The narrow circumstellar \ha\ permitted us to estimate the acceleration efficiency of cosmic rays.
The found value is close to the efficiency inferred in the same way for other two SNe~IIn,
  SN~1997eg and SN~2002ic.
The efficiency of cosmic ray acceleration is utilized to reproduce the radio flux 
  from SN~2008iy for the amplified magnetic field consistent with the saturated 
  turbulent magnetic field in the diffusive shock acceleration mechanism.

\end{abstract}

\section{Introduction} 
\label{sec:intro}

Type IIn supernovae are selected by the presence of narrow circumstellar lines in their spectra 
  \citep{Schlegel_1990}.
They generally reveal strong late time optical luminosity \citep{Filippenko_1991, Stathakis_1991} 
  powered by the ejecta interaction with the dense circumstellar (CS) matter (CSM) 
   \citep{Chugai_1990}.
 This family is highly heterogeneous and includes at least two varieties: 
   (i) SNe~IIn interacting with a dense CS wind, e.g., SN~1988Z \citep{Stathakis_1991}    
   and (ii) less numerous, SNe~IIn 
    interacting with a massive CS envelope presumably lost by a violent pulse ejection,  e.g., 
   SN~2006gy \citep{Woosley_2007,Smith_2007}.
  Noteworthy, SNe~IIP and SNe~IIL also show in early spectra (days-weeks) narrow CS emission lines 
     \citep{Khazov_2016,Fassia_2001} although it would be impractical to classify them as SNe~IIn.
  Yet the physics behind  narrow lines in early SNe~IIP and SNe~IIL has a lot 
    in common with that of SNe~IIn.
    
 Not only the origin of SNe~IIn is poorly understood, the physics of spectra  
  formation in most cases is also far from clear.
The interpretation of the optical phenomena of SNe~IIn in terms of spherically-symmetric model of 
  SN/CSM interaction faces serious problems.
 The point is that emission lines  
  in the spherical models are related to the swept-up shell between the forward and reverse 
  shocks or/and outer unshocked ejecta, so predicted profiles 
  are {\em broad and top-hat}, whereas observed profiles in most cases are more narrow and have 
  "Gaussian" appearence. 
Supernova SN~1986J was first to demonstrate apparent mismatch between the high average 
  expansion velocity of $1.3\times10^4$\kms\ of the radioshell \citep{Weiler_1990}
  and the narrow \ha\ emission with FWHM of $\sim 1000$\kms\ \citep{Rupen_1987}.
The paradox is resolved under assumption that the \ha\ originates from shocked CS clouds   
   \citep{Chugai_1993}.
 Archetypical SN~1988Z with  narrow, broad and intermediate \ha\ components 
   \citep{Filippenko_1991, Stathakis_1991} provides further evidence that the strong 
   \ha\ intermediate component originates from shocked CS clouds \citep{Chugai_1994}.
    
 Of particular significance for the SNe~IIn understanding are events that are well observed 
  photometrically and 
  spectroscopically and discovered close to the explosion moment.    
This is the case of the luminous SN~2008iy that is prominent for the delayed 400-day rise 
  towards broad light maximum and for the tremendous \ha\ luminosity of $\approx7\times10^{41}$\ergs\  \citep{Miller_2010}, only 30\% lower than that of the record-holder SN~2010jl \citep{Zhang_2012}.
The photometric and spectral data of SN~2008iy have been interpreted in terms 
   of the SN ejecta interaction with a dense clumpy wind lost by preSN a century
   before the explosion \citep{Miller_2010}. 
   
\citet{Miller_2010} provide us with a guiding general picture of 
  the phenomenon of SN~2008iy, yet some key issues escaped consideration.
Of top interest are the total mass lost by preSN and the SN explosion energy.
The similarity between SN~2008iy and SN~1988Z emphasised by \cite{Miller_2010} 
  suggests that the \ha\ of SN~2008iy might originate from shocked CS clouds.
The modelling of \ha\ in this scenario would permit us to look at the profile asymmetry 
  and check, whether a dust does form in the CDS and/or ejecta likewise in 
  SN~2010jl \citep{Chugai_2018}.
Although the origin of the \ha\ emission  from shocked clouds has no sensible alternative, 
 one should emphasise the qualitative nature of this picture that leaves open some important issues.
Particularly, a striking smoothenesss of the \ha\ profile in SN~2008iy 
   \citep[cf.][]{Miller_2010} raises a question, whether a random ensemble of shocked 
  clouds is able to produce that profile?
  
The key implicit ingredient of the scenario with shocked clouds is their fragmentation and
  acceleration of fragments \citep{Klein_1994,Klein_2003}, which results in the formation of a 
  broad velocity spectrum of \ha-emitting gas constrained by 
   the \ha\ profile \citep[e.g.][]{Chugai_2018}.
 Given some arbitrariness of the described model one may ask ourselves, 
   whether there is any independent observational evidence for the fragmentation of 
   shocked CS clouds?  
 It could be, e.g., a demonstration that we "see" significantly smaller line-emitting 
    fragments compared to the size of original CS clouds.
 In this regard the O\,I 8446\,\AA\ emission line in SN~2008iy spectra \citep{Miller_2010} 
  may turn out an appropriate tool for probing small scale fragments. 
The point is that, if O\,I 8446\,\AA\ line is of the fluorescent origin, which seems 
    to be true, then we would be able to infer  the \ha\ optical depth.
The latter in turn depends on the size of \ha-emitting fragments.
  
The high resolution Keck I/LRIS spectrum on day 711 after discovery shows narrow CS \ha\  
    with a P Cyg profile \citep{Miller_2010}.
Qualitative analysis of this profile provides some hint that we may be see in this line 
  the specific effect  already met in the CS \ha\ of SN~1997eg and attributed to the preshock 
  acceleration of the CS gas by the cosmic ray (CR) precursor \citep{Chugai_2019}.
  
The outlined issues strongly motivate us to revisit SN~2008iy and to look closer at 
   at these interesting aspects of the phenomenon. 
The paper starts with the light curve modelling  
  that provides us with the CS density distribution, the CSM mass, and the explosion energy.
Thereafter I explore in detail the \ha\ emission including 
  the origin of weak profile asymmetry, problem of profile smoothness, and 
  parameters of line-emitting shocked clouds. 
The O\,I 8446\,\AA\ emission will be used then to infer a size of the line-emitting 
 fragments of shocked clouds.
Finaly, the narrow CS \ha\ is analysed with important implications for the CR acceleration 
  efficiency in SNe~IIn and interpretation of the detected radio \citep{Chandra_2009}.

 
\begin{table*}
\centering
\caption[]{Parameters of light curve models and output values on day 702}
\begin{tabular}{p{1.0cm}|p{1.0cm}|p{1.0cm}|p{1.2cm}|p{1.2cm}|p{1.2cm}|p{1.4cm}|p{1.2cm}}
\hline
 $M$   & $E$    & $M_{cs}^{\dag}$    &  $v_{cds}$  &  $v_{sn}$  &
  $r_{cds}$  & $\rho_0$ & $M_{cds}$ \\
     $M_{\odot}$ & $10^{51}$\,erg & $M_{\odot}$ &  \kms\     &  \kms\   &
 $10^{16}$\,cm & $10^{-17}$\gcmq\ &  $M_{\odot}$ \\ 
\hline
  8  &  3.2    &  11.7    & 2700    & 5200    & 3.1     & 9.3  &   11 \\
  5  &  3.2    &  12.7    & 2500    & 5700     & 3.6     & 10  &   11.6 \\
 \hline
 \parbox[]{6cm}{\small $^{\dag}$ Inside the radius of $4\times10^{16}$ cm.}
 \end{tabular}
\label{tab:param}
\end{table*} 

 \section{CS interaction model} 
\label{sec:lcurve}

\subsection{Bolometric luminosity}
\label{sec:bol}

To reconstruct the bolometric light curve  I rely on the light curve presented by  $i$, and $I$ magnitudes 
   \citep{Miller_2010} with  two initial points in visual band and late three points in $r$ band.
\cite{Miller_2010} find that a bolometric correction can reach -1.2 for the $I$ band. 
Here I adopt BC = -1 mag for all the bands and recover the bolometric light curve with zero 
  extinction correction, in line with \cite{Miller_2010}.    
The bolometric light curve is crucial for the modelling of the CS interaction and 
  independent evidence in favour of the adopted bolometric correction is needed.
  
A strong support of this choice is provided by the high \ha\ luminosity.  
On day 652 from the discovery ($\sim 702$\,d after explosion)  
  the bolometric luminosity recovered for BC = -1 is $L_{bol} \approx 1.2\times10^{43}$\ergs.
The bolometric luminosity is expected to be very close to the unabsorbed X-ray luminosity of shocks. 
On the other hand, this X-ray luminosity can be recovered based on 
   the \ha\ luminosity at this age $L(\mbox{H}\alpha)=6.1\times10^{41}$\ergs\  \citep{Miller_2010}.
This can be done as follows.   
   
The \ha\ emission is powered by the absorbed X-rays.
The deposited energy ($D$\ergs\,cm$^{-3}$) in the form of fast photoelectrons is shared between 
  the Coulomb heating, hydrogen ionization and excitation. 
With the energy fraction spent on the Coulomb heating $f_h$ the   
  rate (cm$^{-3}$\,s$^{-1}$) of the hydrogen ionization and exitation of levels $n > 2$ is approximated by the 
   expression $(D/w)(1 - f_h)$, where $w=22$\,eV suggested by the expression of \citet{Xu_1992}.
Assuming that all the recombinations and excitations on levels $n > 2$ end up with the 
  \ha\ emission one gets the \ha\ emissivity (erg\,cm$^{-3}$\,s$^{-1}$)
\begin{equation}
\epsilon_{32} = (E_{23}/w)(1-f_h)D = \psi D\,,
\label{eq:emis}  
\end{equation}
where $E_{23}=1.89$\,eV is the energy of \ha\ photon. 
The numerical computations of $f_h(x)$ dependence on the ionization fraction \citep{Xu_1992} 
 can be approximated as $f_h=x^{0.28}$ with the accuracy better than 10\% in the range 
  $0.01 \lesssim x < 1$.
For $0.05 < x < 0.1$ (cf. Section \ref{sec:oxy}) $0.43 < f_h < 0.52$ and thus the \ha\ conversion efficiency  $\psi \sim 0.044$.  
With this value the expected absorbed X-ray luminosity should be $\approx 1.35\times10^{43}$\ergs.
This is close to the bolometric luminosity $\approx 1.2\times10^{43}$\ergs\ assuming BC = -1 mag,
  which thus supports the adopted choice for the bolometric correction.

 %
 \begin{figure}
   	\includegraphics[trim= 40 120 0 0,width=1\columnwidth]{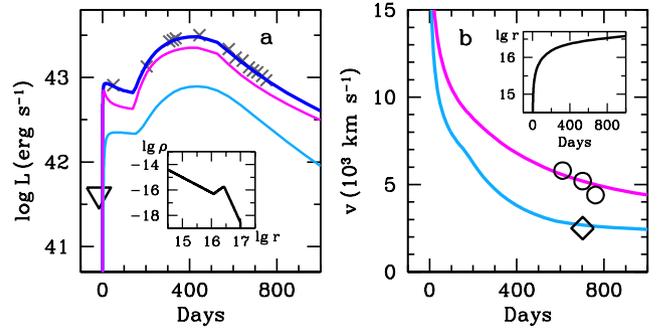}
 	\caption{%
 	Left panel ({\bf a}): bolometric light curve of SN~2008iy ({\em crosses}) and non-detection 
 	upper limit ({\em triangle}) with the model 
 	light curve for 8\msun\ ejecta ({\em blue}) and contributions of the forward shock
 	({\em magenta}) and reverse shock ({\em skyblue}). {\em Inset} shows the CS density.
  	Right panel ({\bf b}): the model CDS velocity ({\em skyblue}) and ejecta velocity at the reverse shock ({\em magenta}); observational maximal ejecta velocity 
 	\citep{Miller_2010} are shown by {\em circles}, while the CDS velocity recovered 
 	from the \ha\ profile  modelling is shown by {\em diamond}.
 	{\em Inset} shows the CDS radius.
 	}
 	\label{fig:lcurve1}
 \end{figure}
 %
  
 %
 \begin{figure}
 	\includegraphics[trim= 40 120 0 0,width=1\columnwidth]{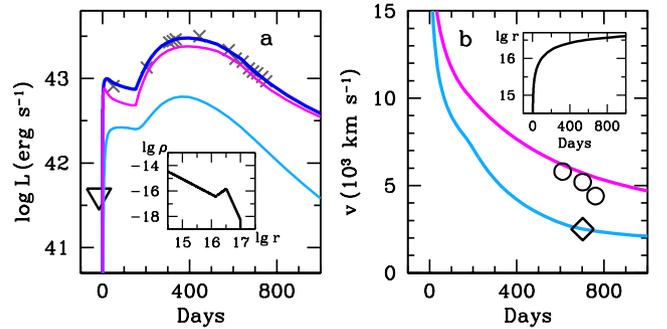}
 	\caption{%
 	The same as Fig. \ref{fig:lcurve1} but for the 5\msun\ ejecta.
 	}
 	\label{fig:lcurve2}
 \end{figure}
 %

\subsection{Modelling interaction}

The exploded preSN could be a WR helium star with the radius of $\sim 10$\,$R_{\odot}$ lost almost all  
 the hydrogen envelope, or the preSN might retain significant amount of hydrogen envelope 
 remaining a red supergiant (RSG) with a typical radius of $\sim 500$\,$R_{\odot}$. 
The different preSN configuration however affects only initial part of the light curve 
  ($t \lesssim 100$ d), so we consider only the case of the compact preSN that reminds us the 
  case of SN~2001em \citep{Chandra_2020}.

It is conceivable that the delayed light curve maximum might be related to 
   the radiation diffusion in the massive CS envelope. 
This however would require a fully ionized 40\msun\ CS envelope. 
The latter in turn would suggest the presence of a narrow CS recombination \ha\ with the luminosity of
  $\sim 10^{43}$\ergs, which is at odds with SN~2008iy spectra. 
The diffusion mechanism for the delayed maximum therefore should be rejected. 
In line with \citet{Miller_2010} the delayed maximum is considered as 
  an outcome of SN ejecta interaction with the CS envelope at about $2\times10^{16}$\,cm.
  
The CS interaction is modelled based on the thin shell approximation   
   \citep[e.g.][]{Chugai_2018} that treats the swept up mass 
   between the reverse and forward shock as a thin shell driven by ejecta dynamical 
   pressure  \citep{Chevalier_1982x}.
Given the instant escape of the generated optical radiation, the shock radiative luminosity 
  is adopted to be equal to the X-ray luminosity of both shocks, which in turn is calculated as 
  the kinetic luminosity multiplied by the radiation efficiency
  $t/(t+t_c)$, where $t$ is the age and $t_c$ is the cooling time of the postshock gas.
The latter is calculated based on the isothermal approximation ($T_e = T_i$) with the postshock 
  density being four times of the preshock density and the cooling function for 
  the solar abundance \citep{Sutherland_1993}.     
The interaction model for the light curve assumes the average smooth CS density 
    set by the broken power law, ignoring clumpiness that is a sound approximation for the probing CS mass and SN energy.
For the explored parameter set both the reverse and forward shocks remain   
  radiative untill day $\sim 700$.  
In reality, however, the forward shock in the intercloud gas is adiabatic.

The initial SN ejecta is set as the homologously expanding envelope ($v = r/t$) with 
   the density distribution $\rho = \rho_0/[1 + (v/v_0)^8]$. 
Parameters $\rho_0$ and $v_0$ are defined via the ejecta mass $M$ and kinetic energy $E$.
The radiation output of the ejecta/CSM interaction is determined by the kinetic energy 
 of the ejecta external layers that can be similar for different combinations of the 
 ejecta mass, kinetic energy, and CS density distribution.
This means that to some degree there is a freedom for the ejecta mass, 
  although constraints from the light curve and expansion velocities and conservative 
   requirements of minimal kinetic energy and CS mass can bound the ejecta mass. 
The general rule is that the larger ejecta mass $> 8$\msun\  requires higher kinetic energy, 
  while lower ejecta mass ($< 8$\msun) requires larger CS mass.
We adopt the ejecta mass of 8\msun\ for the fiducial model, but consider the case of 
 5\msun\ ejecta as well.

Results for model A (8\msun) and model B (5\msun), assuming the explosion occured 50 \,d 
 prior to the detection, are shown in Fig. \ref{fig:lcurve1}  
  and  Fig. \ref{fig:lcurve2}, respectively, (for parameters see Table~\ref{tab:param}).
The Table includes SN ejecta mass,
  kinetic energy, and the mass of the CSM in the range $r < 4\times10^{16}$ cm, 
  and output values 
  at the age $t = 702$\,d: the CDS velocity,  ejecta velocity at the reverse shock, 
  CDS radius, preshock CS density, and the CDS mass (the total swept-up mass is 
  by $\sim 1$\msun\ larger). 
The $^{56}$Ni mass in both models is 0.07\msun\ adopted to be similar to that of     
  SN~1987A; the $^{56}$Ni mass  affects only the light curve minimum at about 100\,d.   
The initial light maximum is related to the diffusion cooling of ejecta 
   calculated based on the \cite{Arnett_1980} approximation.
Note that the forward shock dominates in the bolometric luminosity 
   over the reverse shock.
{ The recovered CS mass and density distribution are well constrained by the light curve 
 and the ejecta velocity.
This is demonstrated by the similar inferred CS mass for different ejecta mass (Table~\ref{tab:param}).  

The major result of the interaction modelling is the large mass of the CS enevelope,  $M_{cs} \sim 10$\msun.
The preshock density on day 702 in the model A of $ 9.3\times10^{-17}$\gcmq\ at the radius of 
  $3.1\times10^{16}$\,cm combined with the wind velocity of 45\kms\ (Section \ref{sec:narrow})
   suggests the tremendous mass loss rate $\dot{M} \approx 0.09$\msun\,yr$^{-1}$ at about 200 yr 
   before the explosion.

 \section{\ha\ and CS clumpiness} 
\label{sec:ha}

\subsection{Background}

The \ha\ emission of SN~2008iy dominated by the intermediate component with 
  the half width at half maximum of $\sim 1000$\kms\ presumably originates from 
  CS clouds shocked in the forward shock likewise in the case of SN~1988Z \citep{Chugai_1994}.
Multi-dimensional computations of the blast wave interaction with an interstellar 
  cloud \citep{Klein_1994,Klein_2003}   
  and laboratory experiments \citep{Klein_2003,Hansen_2007} suggest the following qualitative 
  picture.
  
The forward shock with the postshock speed $v_{ps} \approx v_{cds}$ and density $\rho_s$ colliding 
   with the CS cloud of the radius $a$ and density $\rho_c$ drives the transmitted (cloud) 
   shock with the speed  $v_c \approx v_{ps}(\rho_s/\rho_c)^{0.5}$. 
The fast forward shock overtakes the cloud in the time $\sim a/v_{ps}$ and 
  interacts with itself 
   creating a backside high pressure that drives also the rear cloud shock. 
The cloud eventually gets crushed by both opposite shocks into the umbrella-shaped pancake 
  that is rolled-up by shear flow into a vortex ring.
The latter undergoes strong disruptive instabilities with the eventual turbulent distruction.   
The cloud crushing time is $t_{cc} \sim a/v_c$ and 
  it takes $t_f \sim 4\times t_{cc}$ for shear flow to roll-up the vertex ring, disrupt it   
    into small fragments, and accelerate them to the speed  
   comparable to $v_{ps}$.
Finally, tiny fragments get mixed with the hot gas of the forward shock.  
The interaction with the clumpy CSM thus results in the two-phase structure 
   of postshock layer with cold dense fragments 
   in a broad range of scales and velocities imbedded into the hot gas.
These cold fragments ionized by X-rays are responsible for the observed Gaussian-like \ha\ 
   of SN~2008iy.   
   
To reproduce the \ha\ emission by direct numerical simulations 
  currently is beyond reach.
The \ha\ formation in SN~2008iy therefore will be  desribed below  
   based on simplified models.  
 
\subsection{Profile modelling}
\label{sec:prof}
 
The model that accounts for the \ha\ profile should include the emission of a random ensemble of 
   shocked clouds and accelerated cloud fragments with some velocity spectrum.
To cope with this complicated situation I assume the constant specific 
  \ha\ emissivity and  adopt the appropriate spherical mass-velocity distribution in the 
  postshock layer between the forward shock and the CDS
\[
g(v) = \left\{
\begin{array}{rl}
  (v/v_1)^2             & \mbox{if $v<v_1$} \\
  (v_2-v)/(v_2-v_1)     & \mbox{if $v_1<v<v_2$}\\
     0                  & \mbox{otherwise}\,
 \end{array} \right.
\]
    where $v_1$  and $v_2$ are parameters determined by the \ha\ profile fit.
 The $v_1$ value is expectedly close to the cloud shock speed $v_c$,  
    whereas the maximal velocity of accelerated cloud fragments $v_2 < v_{cds}$. 
The low velocities range $v < v_1$ qualitatively takes into account the emission from 
  the cloud that is not yet fragmented and the vortex ring generation by the KH instability. 
The rotational velocities of the vortex combined with radial 
   velocities result in the emegence of stripped cloud material with 
   low velocities $v < v_1$ in the frame of the SN center.
   
Currently 3D-hydrodynamic simulations of the shock-cloud interaction are performed 
	only for the adiabatic case.
There is no, therefore, a relevant numerical example of the velocity spectrum for the ensemble of shocked CS clouds in the forward shock for the scenario we consider.	
The qualitative explanation of how this velocity spectrum in the range $v > v_1$ does 
   form can be given in terms of the 
    constant rate of CS clouds inflow into the forward shock,
  the cloud lifetime $t_0$, the cloud survival probability $p(t) \propto (1 - t/t_0)$, 
   and the steady acceleration of cloud fragments by the forward shock flow  \citep{Chugai_2018}.   
     
 The radial coordinate of clouds and their fragments is assumed to be a random value 
    in the shell $\Delta r$ between the CDS and the forward shock.
 The $\Delta r$  value is not defined in the thin shell model.
 One cannot apply also the self-similar solution for the driven adiabatic shock 
   in which case  $\Delta r$ is larger than for the radiative shock.
  We adopt $\Delta r/r_{cds} = 0.2$ compared, e.g., to 
  $\Delta r/r_{cds} = 0.27$ in the adiabatic self-similar solution for the 
  homologously expanding  ejecta ($\rho \propto v^{-8}$) in the steady wind 
  ($\rho \propto r^{-2}$)  \citep{Chevalier_1982s}.
 The effect of the $\Delta r$ choice is explored a bit below.
   
 On day 652 after discovery (post-explosion day 702) the \ha\ and the model 
    calculated by the Monte Carlo (MC) technique are shown in Figure \ref{fig:ha}.
The appropriate parameters of the velocity spectrum $g(v)$ are $v_1 = 950$\kms\ and  $v_2 = 2300$\kms.     
 Shown are three versions:  (a) without any continua absorption or scattering, 
 (b) with the CDS optical depth $\tau_{cds} = 0.2$ produced possibly by dust or other 
  absorbing agent; (c) with Thomson scattering in the unshocked ejecta 
    ($\tau_{sn} = 0.25$) assuming homogeneous distribution of electron number density. 
 The case with the Thomson scattering is preferred compared to other versions. 
 We checked effect of zero albedo and found that the difference with the Thomson 
  scattering is negligible, so formally the dust in ejecta is not rulled out. 
 A possible additional contribution of the CDS combined with the 
  Thomson scattering should be small,  $\tau_{cds} < 0.1$.
The uncertainty of the $\Delta r$ choice affects only the optimal value of the 
  optical depth of unshocked ejecta. 
For  $\Delta r/r_{cds} = 0.15$ and 0.25 the ejecta optical depth is 0.2 and 0.3, respectively. 

In the optimal case (c) the \ha\ is fully dominated by the emission of 
  shocked clouds with only 1.6\% contribution from unshocked ejecta, consistent 
   with the Thomson optical depth, and 10\% contribution from the CDS emission.
In the latter case we admit the boxy \ha\ profile originated from CDS fragments produced 
  by the Rayleigh-Taylor instability of the CDS \citep[e.g.][]{BloEll_2001}.   
 The \ha\ luminosity on day 702 is of $6.1\times10^{41}$\ergs\ \citep{Miller_2010} 
   of which $5.4\times10^{41}$\ergs\ (i.e. 88\%) is related to the shocked CS clouds.
   
The weak \ha\ blueshift suggests that the significant amount of dust is not  
  formed at the considered age (702 d) neither in CDS nor in the ejecta.  
Basically, the dust is expected to form at $t > 300$ days in the CDS \citep{Pozzo_2004} or ejecta. 
For example, in the well observed SN~IIn 2010jl at the age 
   $t \gtrsim 400$\,d  the presence of the dust in ejecta and  CDS is 
   indicated by the significant \ha\ blueshift \citep{Maeda_2013,Chugai_2018}.
The dust absence in SN~2008iy on day 702 could be related to the unusually 
  high late time luminosity and, therefore, high 
  gas temperature that inhibits the dust formation.
Indeed, with the CDS radius  of $3.1\times10^{16}$\,cm 
    and the luminosity of $10^{43}$\ergs\  the effective temperature at this age 
    is $\approx 2000$\,K. 
This value slightly exceeds the condensation temperature of the graphite (1800\,K)      
  \citep[e.g.][]{Kozasa_1989} that is the maximal among astrophysical dust species.  
  
Yet the CDS with the column density of 2\,g\,cm$^{-2}$ and the temperature of 
  $\sim 2000$\,K could contain significant amount of H$^{-}$ molecules, which are  
   efficient absorbers in the \ha\ band.    
To estimate the optical depth produced by H$^{-}$ in the model A we use the CDS density 
  suggested by the upstrean ram pressure 
  $\rho = \rho_0(v_s/c_s)^2 = 2.4\times10^{-11}$\gcmq,  where $c_s \approx 5.2$\kms\ is the 
  sound speed of the CDS gas. 
We adopt the presense of 1\msun\ of hydrogen in ejecta, whereas in the CSM 
  the hydrogen abundance is assumed to be 0.7.
The electron number density is calculated for the absorbed X-ray luminosity of $10^{43}$\ergs\
  deposited in the CDS on day 700, while the number density of H$^{-}$ is determined by 
  the Saha equation.
The inferred CDS optical depth due to H$^{-}$ is $\tau_{cds} = 0.08$ that is marginally  
   consistent with the upper limit ($<0.1$) imposed by the \ha\ modelling.

 %
 \begin{figure}
 	\includegraphics[trim= 50 60 0 0,width=1\columnwidth]{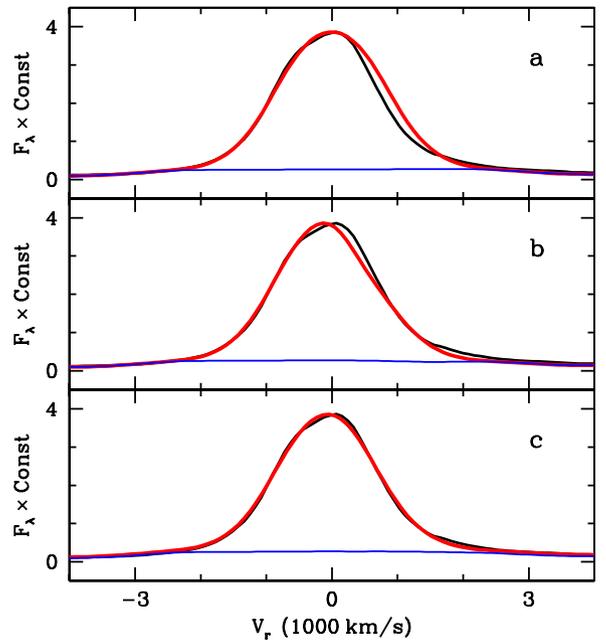}
 	\caption{%
 	The \ha\ line on day 652 after discovery ({\em black}) with model profiles: {\bf a} -- 
 	without any continuum absorption/scattering,  {\bf b} -- with the continuum absorption in 
 	the CDS, {\bf c} -- with the Thomson scattering in the unshocked 
 	ejecta.
 		}
 	\label{fig:ha}
 \end{figure}
 %

 \subsection{\ha-emitting clouds and fragments} 
\label{sec:smooth} 

\subsubsection{Cloud size and density}  
 \label{sec:filf} 
 
The undisturbed cloud density $\rho_c$ is related to the average preshock CS density 
  $\rho_0$, cloud filling factor of undisturbed CS clouds $f_0$, and the 
    intercloud mass fraction  $\phi$ as  
\begin{equation}
\rho_c  = \rho_0(1-\phi)f_0^{-1}\,,
\label{eq:filf}
\end{equation}
In turn, the cloud shock with the velocity $v_c \approx 900$\kms\ is driven by the postshock flow 
with the density  $\sim 4\rho_0$ and velocity $v_{ps}= 2700$\kms\ on day 702, so that 
  $v_c=v_{ps}(4\rho_0/\rho_c)^{1/2}$ and therefore $\rho_c/\rho_0 =  4(v_{ps}/v_c)^2 \approx 36$.  
Note, for the forward postshock density we use the average, and not intercloud density. Doing in this way we take into account fragmentation and mixing of tiny fragments of shocked clouds with the shocked intercloud gas.   
Following the similar case of SN~1997eg \citep{Chugai_2019} we  adopt here $\phi = 0.5$. 
These values along with the Equation (\ref{eq:filf}) result in $f_0 = 0.014$.   
For the average density $\rho_0 \approx 9\times10^{-17}$\gcmq\ on day 702
   (Table \ref{tab:param}) the unperturbed cloud density is thus
   $\rho_c = 3\times10^{-15}$\gcmq\ 
   with the hydrogen number density $n_c = 1.3\times10^9$\cmq\ assuming the hydrogen abundance 
   X = 0.7, while for $\phi = 0.5$ the intercloud density is $n_{ic} = n_cf_0/(1 - f_0) = 2.5\times10^7$\cmq.
   
The cooling time of the cloud shock is  
\begin{equation}
t_{cool} \approx \frac{m_pv_c^2}{32n_c\Lambda} \approx 10^4\, \mbox{s}\,,
\end{equation}
where the cooling function $\Lambda=3\times10^{-23}$\,erg\,s$^{-1}$\,cm$^3$ for $T=10^7$\,K 
  \citep{Sutherland_1993}.
The condition that the cooling time should be less than the cloud crushing time $t_{cc} = a/v_c$ 
  implies the lower limit of the cloud radius $a > v_ct_{cool} \approx 10^{12}$\,cm.
  
The upper limit of $a$ is determined by the condition that the cloud fragmentation time 
  $t_f \sim 4t_{cc}$ should be less than the time it takes for the forward shock to advance by 
  the distance $\Delta r = 0.2r_{cds}$, i.e., $t_f < \Delta r/v_{ps}$, which 
  results in $a < (\Delta r/4)(v_c/v_{ps}) \approx 5\times10^{14}$\,cm.  
This upper limit  imposes a constraint on the number of CS clouds ($N_c$) 
  in the layer between the forward shock and 
  the CDS from the obvious equality $(4\pi/3)a^3N_c = f_0V$, where $V = 9\times10^{49}$\,cm$^3$ 
  is the layer volume.
The inequality $a < 5\times10^{14}$\,cm then implies the lower limit 
 $N_c > 10^3$.

The recombination \ha\ luminosity from shocked clouds and fragments in the forward shock layer 
  $\Delta r = 0.2r_{cds}$  with the volume $V = 9\times10^{49}$\,cm$^3$ on day 702 
   is  $L_{32} = \alpha_{32}n_e^2VfE_{23}$, 
  where $f \ll f_0$ is the filling factor of the line-emitting fragments,
  $E_{23}$ is \ha\ photon energy, $\alpha_{32} = 1.8\times10^{-13}$\,cm$^3$\,s$^{-1}$ 
   is the \ha\ effective recombination coefficient in the recombination case C (opaque Balmer lines)  
   for $T_e = 10^4$\,K.
For the \ha\ luminosity of the shocked clouds on day 702
  of $5.4\times10^{41}$\ergs\ the electron number density is therefore 
   $n_e = 8\times10^8\eta^{1/2}$ \cmq\, where $\eta = f_0/f$ is the compression factor 
   that exceeds adiabatic value ($\eta = 4$) due to the radiative cooling and isobaric 
   compression of the shocked gas.
The upper limit for the compression factor in the radiative cloud shock is  
  $\eta = (v_c/c_s)^2 \sim 10^4$, where $c_s \sim 10$\kms\ is the sound speed in 
  \ha-emitting gas.
However the real value of $\eta$ can be lower and will be estimated in Section \ref{sec:oxy}.

 %
 \begin{figure}
 	\includegraphics[trim= 40 60 0 0,width=0.95\columnwidth]{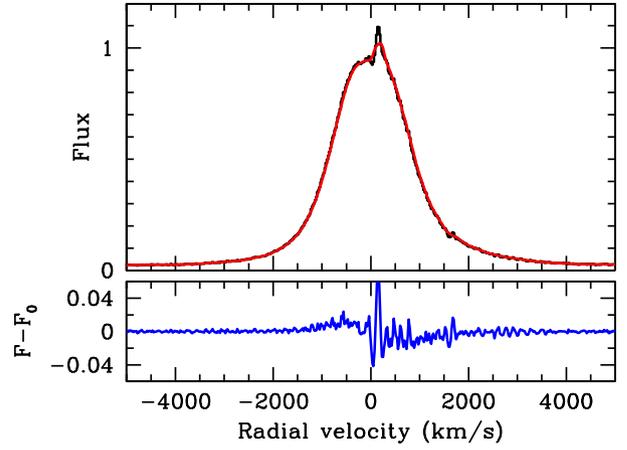}
 	\caption{%
 	\ha\ in SN~2008 spectrum on day 711 after discovery ({\em black}) with overploted smoothed version. 
 	The lower panel shows residual between spectrum and smoothed version to demonstrate the 
 	fluctuations related to the clumpiness of the line-emitting gas. Fluctuation amplitudes
 	ralated to clumpiness (if any) do not exceed 1\%.
 	Narrow CS \ha\ produces in the residual prominent feature at zero velocity.
 	The observational profile is shifted toward the red by 120\kms\ in order to compensate the blueward 
 	skewing related to the ejecta scattering (absorption).
  	}
 	\label{fig:resid}
 \end{figure}
 %


\subsubsection{\ha\ smootheness}

The notable smootheness of the \ha\ line on day 711 (Fig. \ref{fig:resid}) with the 
  flux fluctuation at the velocity scale of $\sim 200$\kms\ of $\lesssim 1$\% 
  is challenging problem for the scenario of clumpy line-emitting zone. 
Obviously, the number of line-emitting fragments of shocked clouds 
  should be large enough to significantly reduce a fluctuation amplitude.
  
The minimal number of cloud fragments imposed by the low amplitude of flux fluctuations 
   can be estimated via MC simulations. 
We consider here the emission of shocked CS clouds omitting small contribution of 
     unshocked ejecta and CDS.
Generally, one has to distiguish between two populations of line-emitting clouds:
   non-fragmented clouds (simply "clouds") with the number of $N_c$ 
   and their numerous fragments ($N_f \gg N_c$). 
Here we consider the simplified set up assuming that both clouds and fragments 
  has the same luminosity weight $w = 1$, while their   
  velocity are diced according to the distribution $g(v)$.   
The cloud position on the sphere is random.  
The resulting profile is then convolved with the Gaussian filter adopting the resolution of  
  100\kms.
  
The effect of fragment number is demonstrated in Figure \ref{fig:cloud} for $N_f = 10^4$ and 
   $10^6$. 
Only in the case of $N_f = 10^6$ fluctuations become comparable to those observed on day 711. 
At first glance this number is unrealisticaly large. 
In fact not, because to provide $10^3$-fold increase of fragments number compared to the 
  shocked clouds ($\sim 10^3$) the cloud of the radius $a$ should break up into cloudlets with the 
  radius of only $0.1\times a$.
As will be shown below (Section \ref{sec:oxy}) the cloud fragmentation cascade spans the 
  scale range exceeding 2.3 dex, which implies that $N_f > 10^6$.
  
Summing up, the smootheness of \ha\ is simply an outcome of the large number of line-emitting 
  cloud fragments ($> 10^6$) in the forward shock.
Noteworthy that the scenario of the line emitting fragments predicts the maximal fluctuation  
 amplitude in the low velocity range ($|v_r| < 1500$\kms) of the profile (Fig. \ref{fig:cloud}), which seems to be consistent with the picture of fluctuations in the observed \ha\  
  (Fig. \ref{fig:resid}).
  
 %
 \begin{figure}
 	\includegraphics[trim= 20 90 0 0,width=0.95\columnwidth]{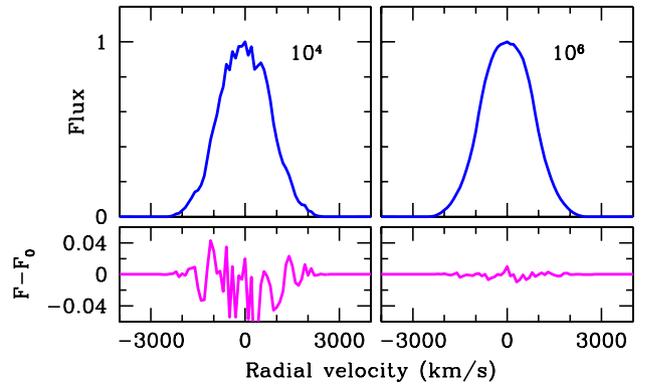}
 	\caption{%
 	Monte Carlo simulation of \ha\ line emission produced by superposition of random line-emitting 
 	fragments of shocked clouds. Shown are cases with $N_f = 10^4$ ({\em left}) and $10^6$ ({\em right}). 
 	Lower panels show residual with respect to the case of $10^8$ fragments.
 	The fluctuations at the level $\lesssim 1$\% require $N_f \gtrsim 10^6$. 
 	}
 	\label{fig:cloud}
 \end{figure}
 %

 \subsection{Hydrogen ionization} 
\label{sec:ioniz}  

 We already saw (Section \ref{sec:bol}) that to account for the \ha\ luminosity at the age 
  of 702\,d the line-emitting gas should absorb X-rays with the luminosity of 
   $\sim 10^{43}$\ergs\ comparble to the bolometric luminosity.
 The hydrogen ionization of shocked CS clouds can be maintained by 
   three sources: (i) X-rays of the reverse and forward shock, (ii) X-rays of cloud shocks, and 
   (iii) XUV radiation from a hot boundary layer at the interface between the cold dense 
   gas of shocked clouds and the hot gas of the forward shock.   
 The X-rays from the reverse shock are significantly absorbed by the CDS in the range 
  $h\nu < 4.5$ keV and given low luminosity of the revese shock it does not contribute 
  significantly to the ionization of shocked clouds.
  
For the X-ray of the forward shock to be efficient ionizing agent of shocked clouds 
 their surface area $S$ should provide efficient covering, i.e., their area 
 ratio $A = S/4\pi r_{cds}^2$ should be greater than unity. 
If this is the case,  the radial hydrogen column density determined by the total
  CS cloud mass $\rho_0V(1-\phi)$ is  $N_{\mbox{\tiny H}} \sim 1.3\times10^{23}$\,cm$^{-2}$.
This value implies the efficient absorption of X-rays in the range $h\nu < 4$\,keV.     

X-rays of cloud shocks with the temperature of $\sim 1$ keV provide the luminosity 
  $L_c = 2\pi C r_{cds}^2\rho_cv_c^3 = 1.5\times10^{43}C$\ergs, where $C$ is the covering factor of shocked clouds, $C = (3/2)(f_0V/a)/(4\pi r_{cds}^2) \gtrsim 0.3$, since 
  $a < 5\times10^{14}$\,cm.
This means that X-ray emission of shocked CS clouds is able to significantly contribute 
 to the hydrogen ionization of shocked CS clouds.
  
The hydrogen of cold fragments could be ionized also via the electron  conductivity flux 
 from the background hot plasma of the forward shock ($T \approx 8$\,keV).
The conductivity heating of cold dense fragments is impeeded by the magnetic field 
  of dense fragments. 
Yet a hybrid mechnism could operate that includes an extreme UV radiation 
  dominanted by He\,II 304\,\AA\ from 
  a hot dense boundary layer at the interface between cold fragments and hot gas of the 
  forward shock.
The heat flux can be written as a fraction $\zeta$ of the saturation flux
   $F_{sat} = 0.2nkT_eu_e$ \citep{Cowie_1977}, where $u_e$ is the electron mean thermal velocity.
For $T_e = 8$ keV and  $n = 4n_{ic} = 10^8$\cmq\ one gets $F_{sat} = 2\times10^9$\ergs\,cm$^{-2}$.
The area of the interface between hot and cold phases is 
 $S \sim 10^{35}(A/10)$\,cm$^2$, which implies the conductive luminosity 
  $L_{con} = \zeta F_{sat}S = 2\times10^{44}(A/10)\zeta$\ergs. 
Even for $\zeta \sim 0.1$, the heat conductivity meets the requirement for the \ha\ energizing.
    
 We thus find ourselves in a situation when all three sources (X-rays from the forward shock, X-rays from  
   shocked clouds, thermal conductivity) can efficiently contribute to the hydrogen ionization 
   of shocked clouds. 
 At the moment we are not able to conclude, which of these mechanisms dominates.
 
 %
 \begin{figure}
 	\includegraphics[trim= 80 120 0 0,width=1\columnwidth]{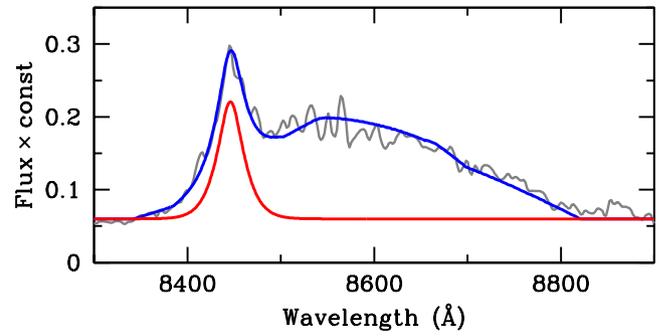}
 	\caption{%
 	O\,I 8446\,\AA\ emission blended with Ca\,II triplet in the SN~2008iy spectrum on 
 	 day 652. {\em Red} line shows deblended contribution of O\,I 8446\,\AA.
 	 {\em Left} panel shows the case of similar line profile for O\,I line and Ca\,II lines, 
 	 which indicates that Ca\,II lines should be broader.
 	{\em Right} panel shows the case of broader Ca\,II lines.
 	{\em Inset} shows the normalized  emissivity distribution for O\,I line ({\em red}) 
 	and Ca\,II lines in this case. 	 
 	}
 	\label{fig:oxy}
 \end{figure}
 %

 \subsection{O\,I 8446\,\AA\ and cloud fragmentation} 
\label{sec:oxy}

\subsubsection{\ha\ optical depth}

Between  days 560 and 652 after discovery SN~2008iy spectra show O\,I 8446\,\AA\ emission line 
   \citep{Miller_2010}  at the blue side of the Ca\,II 8600\,\AA\ triplet. 
The absence of the similar O\,I 7774\,\AA\ emission indicates that the 8446\AA\ emission 
   cannot be produced by the oxygen recombination or collisional excitation in which case 
   the intensity of both lines whould be comparable \citep[cf.][]{Borkowski_1990}. 
The likely origin of O\,I 8446\,\AA\ is therefore the fluorescence \citep{Bowen_1947} due to the 
   absorption of L$\beta$ 1025.728\,\AA\ in the ultraviolet transition 
   2p$^3$P - 3d$^3$D$^{\circ}$  of O\,I 1025.762\,\AA.
Noteworthy, \cite{Leibundgut_1991} have identified  O\,I 8446\,\AA\ emission in the 
  spectrum of SN~1986J and suggested its fluorescent origin.
   
The O\,I 8446\,\AA\ line profile on day 652 recovered by deblending 
  (Fig. \ref{fig:oxy})  suggests the flux ratio $R = F(8446)/F(\mbox{H}\alpha) = 0.05$.
This ratio depends on the \ha\ optical depth $\tau_{23}$ \citep{Netzer_1976} and thus 
  can be used to estimate the hydrogen excitation degree.
Note the obvious difference of the profiles of O\,I and Ca\,II lines (Fig. \ref{fig:oxy}), 
  which reflects a different origin of fluorescent O\,I line and collisionally 
   excited Ca\,II lines.

The L$\beta$ photon emitted with the branching ratio $p_1 = 0.56$ can be 
  absorbed by the oxygen with the probability $p_2$ that depends on the oscilator strength of 
  corresponding H and O\,I transitions and oxygen abundance.
Note that the hydrogen and oxygen are mostly neutral in the shocked clouds where \ha\ and 
  O\,I 8446\,\AA\ presumably originate from. 
Assuming the solar O/H ratio ($6\times10^{-4}$ by number) and using NIST data for oscilator  
  strengths of both transitions one finds $p_2 = 8.4\times10^{-5}$.
The L$\beta$ absorption by the oxygen is followed by the   
  cascade of 11287\,\AA\ and 8446\,\AA\ lines with the branching ratio $p_3 = 0.3$. 
The exitation of the third hydrogen level thus ends up by the 
  emission of O\,I 8446\,\AA\ photon with the probability 
  $p_{31} = p_1p_2p_3 = 1.41\times10^{-5}$.
The flux ratio of O\,I 8446\,\AA\ and H$\alpha$ is then
\begin{equation}
R =  \frac{A_{31}p_{31}}{A_{32}\beta_{23}}\left(\frac{6563}{8446}\right) = 
 1.38\times10^{-5}\beta_{23}^{-1}\,,
 \label{eq:ratio}
\end{equation}
where $\beta_{23}$ is the \ha\ escape probability. 
For the observational value $R = 0.05$ the Equation (\ref{eq:ratio}) gives    
  $\beta_{23} = 2.8\times10^{-4}$ or the average number of \ha\ photon scattering 
   $N_s = 1/\beta_{23} = 3.6\times10^3$.
   
 The fragments of the dirupted cloud have an intricate geometry \citep[e.g.,][]{Klein_2003} 
   that can be approximately viewed as a combination of sheets and filaments. 
 The cumulative surface $S$ of fragments is described by their number $N$, the 
   minimal $h$ and maximal $l$ scales of typical fragment as $S \sim Nl^2$ for sheets and 
    $S \sim Nlh$  for filaments.
 Noteworthy that the expression for the total volume of sheets and flaments is the same 
    $Sh = fV$, i.e., independent of the fragments geometry.
 This expression establishes the useful relation between $h$ and $S$, or area ratio $A$ for a    
   given $f$ and CDS radius $r_{cds}$.
 
To estimate the \ha\  optical depth from the number of scatterings we approximate 
  a typical fragment by a plane slab with the central \ha\ optical depth $\tau_{23}$.
For the homogeneous density and excitation rate in the slab the \ha\ raiation transfer 
   with the Doppler absorption coefficient and the    
   complete frequency redistribution results in the average number of scattering 
\begin{equation}
N_s = Z\tau_{23}\sqrt{\pi \ln{\tau_{23}}}\,,
\end{equation}
with $Z\sim 1$  \citep{Capriotti_1965}.
Our direct MC simulation of double diffusion (space and frequency) results in $Z = 0.46$ in the 
  range of $1000 < \tau_{23} < 2000$.
Given the observational estimate of $N_s$ one gets then $\tau_{23} = 1.6\times10^3$.

\subsubsection{Evidence for cloud fragmentation}

The \ha\ optical depth combined with the luminosity of \ha\ emitted by shocked clouds 
  permit us to determine two parameters related to the \ha-emitting fragments of 
  shocked clouds -- compression factor $\eta$ and the minimal size $h$. 
This can be done via modelling the hydrogen ionization and excitation.

Equations of the statistcal equilibrium for two-level plus
  continuum hydrogen atom are solved for the electron number density 
  $n_e = 8\times10^8\eta^{1/2}$  
  suggested by the \ha\ luminosity on day 702, the hydrogen number density 
  $n = 1.3\times10^9\eta$ 
  at the same age, and the deposition rate of the absorbed X-rays  
  $D = L(\mbox{H}\alpha)/(\psi fV)$, imposed by \ha\ luminosity 
   $L(\mbox{H}\alpha = 5.4\times10^{41}$\ergs, with the conversion efficiency 
   $\psi$ defined  by the Equation (\ref{eq:emis}), the filling factor $f = f_0/\eta$, 
  and the volume of the forward shock layer $V = 9\times10^{49}$\,cm$^3$.
 Apart from the non-thermal excitation and ionization the model includes 
   collisional transitions, recombinations on levels $n \geq 2$ that eventually populate 
   the second level, the two-photon decay, and  L$\alpha$ scattering. 

The statistcal equilibrium is calculated in the temperature range $T_e \geq 7000$\,K.
  The law limit is set based on the fact that the cooling function for $T_e = 7000$\, is 
  $3\times10^{-25}$\ergs\,cm$^3$, i.e. by 2\,dex lower than at the cloud shock, and rapidly 
   drops for the decreasing temperature \citep{Judge_1990}.
 In fact, even 7000\,K is already unattainable, but we include this value for generality.  
 Constraints imposed by the \ha\ optical depth $\tau_{23} = 1.6\times10^3$ and 
   \ha\ luminosity suggest solutions for $T_e = 7000$\,K, 8000\,K, and 9000\,K as follows 
     $(\eta, A, x)=(51,400, 0.09)$, (58, 450, 0.08), and (460, 800, 0.03).
  Taking into account that $f = f_0/\eta$ one gets for $T_e > 7000$\,K the 
  minimal length scale of fragments $h = fV/(4\pi r_{cds}^2A) < 5\times10^9$\,cm. 

The found size of line-emitting fragments is by more than 2.3\,dex lower compared to 
  the lower limit of the cloud radius $10^{12}$\,cm inferred 
   from the  condition of fast cooling in the cloud shock (Section \ref{sec:smooth}). 
This is the direct outcome of the fragmentation cascade 
  accompanying the disruption of the shocked CS clouds in the forward shock. 
It is easily to show that for the Kolmogorov turbulence the complete fragmentation cascade 
  takes the time comparable to the fragmentation time of the largest  
  scale, which in our case is of the order of the cloud crushing time $t_{cc} \sim a/v_c$.
 %
 \begin{figure}
 	\includegraphics[trim= 40 120 0 0,width=0.9\columnwidth]{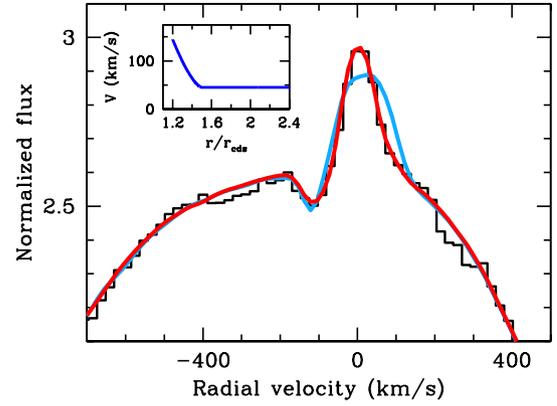}
 	\caption{%
 	Narrow \ha\ in the Keck I spectrum on day 711 after discovery ({\em black}) with the overplotted optimal model ({\em red}). The model suggests the preaccelerated CSM with the speed at the forward shock of 145\kms. 
 	{\em Inset} shows the velocity of the CSM with the preaccelerated inner layer.
 	The alternative model with the constant wind speed of 120\kms\ ({\em skyblue}) is able to describe the absorption component but cannot reproduce the emission component.
 	 	   	}
 	\label{fig:nar}
 \end{figure}
 %

\section{Narrow \ha\ and CR acceleration}
\label{sec:narrow}

\subsubsection{Modelling narrow \ha}

The high-resolution Keck spectrum on day 711 day after discovery
  shows, on top of the intermediate component, the narrow \ha\ with the P Cyg profile that 
  apart from the scattering contains also the net emission 
  with the luminosity of $(2.8\pm0.6)\times10^{39}$\ergs\ \citep{Miller_2010}. 
In the model A (8\msun\ ejecta) on day 761 the volume emission measure of the CSM outside the shock 
  ($> 3.3\times10^{16}$\,cm) in terms of smooth density and full ionization is 
  $EM = 1.5\times10^{64}$\,cm$^{-3}$. 
With the \ha\ effective recombination coefficient for $T_e = 10000$\,K  of 
   $1.18\times10^{-13}$\,s$^{-1}$\,cm$^3$  
   \citep{Martin_1988} one gets the luminosity $L(\mbox{H}\alpha) \approx 5.3\times10^{39}$\ergs.
Although simple in approach, this estimate indicates that the refined model with the \ha\  
   originated  
  from partially ionized clouds could be closer to the observational luminosity of narrow \ha.
Below we model the narrow \ha\ in terms of a relative flux.   
 
First, one should to pinpoint the rest frame.
Narrow CS emission lines of [O\,I] 6300.30, 6363.78\,\AA\ doublet 
  in the spectrum on day 711 show additional redshift of $+25\pm5$\kms
  compared to $z = 0.0411$. 
This small correction is implemented in the rest frame wavelength in Figure \ref{fig:nar}; 
  the excellent match between the observed emission peak and the model supports the  
  correction. 
  
The modelling of the narrow \ha\ indicates that the observed profile cannot be reproduced 
  assuming the constant wind velocity along the radius that 
 is demonstrated for the case of the constant CSM velocity of 120\kms\ (Fig. \ref{fig:nar}). 
While the absorption component can be described in this scenario, the emission component 
  turnes out to be unacceptably broad. 
The appropriate model (Fig. \ref{fig:nar}) includes the outer wind with the velocity of    
  45\kms\ and the preshock accelerated wind with the velocity decreasing outward  
   from 145\kms\ at the shock ($r_s = 1.2r_{cds}$) down to 45\kms\ at the radius 
   $r = 1.5r_{cds}$.
The velocity of the accelerated wind is constrained by the blue absorption wing, 
  whereas the outer wind velocity is constrained by the narrow upper part of the 
  CS emission component.
The Keck spectrum on day 711 shows narrow coronal 
  line [Fe\,X] 6374\,\AA\ \citep{Miller_2010} that presumably  
  originates from the hot intercloud medium photoionized by X-rays of the forward shock 
  likewise in SN~1997eg \citep{Chugai_2019}.
The narrow \ha\ therefore likely originates from the cloudy component of the CSM.  
The depth of the absorption component suggests that the clouds filling factor 
   $f_c \geq 0.05$; the case $f_c = 0.05$ is shown in Figure \ref{fig:nar}.
This value exceeds the earlier estimate $f_0 = 0.014$ (Section \ref{sec:filf}).
The contradiction can be reconciled, if clouds have a core and a rarefied halo with the radius 
  by a factor of 1.5 larger than that of the core. 

Noteworthy, to reproduce the asymmetry of the broad (intermediate) component on day 761
  (Fig. \ref{fig:nar}) the continuum optical depth of the CDS combined with the unshocked ejects should be 0.55, compared to the optical depth 0.25 on day 702.
This indicates that the dust probably starts to form. 
The conjecture is supported by the effective temperature at this stage 
 (1700\,K) that is slightly lower than the graphite condensation temperature (1800\,K).
 
 \subsubsection{CR precursor}
 
 The preshock CSM acceleration implied by the narrow \ha\ 
   has been invoked earlier to account for the similar CS \ha\ in SN~1997eg (IIn) on 
   day 198 in which case it is atributed to the CR precursor \citep{Chugai_2019}.
The CR are presumably accelerated by 
   the standard mechanism of the diffusive shock acceleration (DSA) 
    \citep{Axford_1977,Krymskii_1977}.  
The CR pressure ($p_c$) at the forward shock with respect to the upstream ram pressure 
 ($\rho_0v_s^2$) can be estimated using the undisturbed CS velocity ($u$), preshock 
  CS velocity ($u_{max}$) and the shock speed ($v_s$) based on the equation of the 
   momentum conservation 
\begin{equation}
\rho v\frac{\partial v}{\partial x} + \frac{\partial p_c}{\partial x} = 0\,,
\end{equation}
   where I neglect the upstream gas pressure.
Along with the mass conservation, the integration in the upstream flow in the shock frame 
  with the boundary conditions 
  $v(-\infty) = v_s-u$, $p_c(-\infty) = 0$ and $v(0) = v_s-u_{max}$, $p_c(0) = p_c$ 
 results in the efficiency of CR acceleration 
\begin{equation}
 \epsilon \equiv p_c/\rho_0 v_s^2 = (u_{max} - u)/v_s\,, 
\label{eq:precur}
\end{equation}
   where small quadratic terms are omitted.
With $u_{max} = 145$\kms, $u = 45$\kms\ and $v_s = 2600$\kms\ at $t = 761$\,d    
   the efficiency of CR acceleration is then $\epsilon = 0.038$.

The late high resolution spectra of \ha\ are available for two other SNe~IIn:  
 SN~1997eg on day 198 \citep{Salamanca_2002} and SN~2002ic on day 256 \citep{Kotak_2004}.
For SN~1997eg the undisturbed CS velocity $u = 20$\kms\ and preshock maximal velocity 
  $u_{max} = 160$\kms\ are infered from the [Fe\,X] 6374\,\AA\ and \ha\ profile modelling, 
  whereas the forward shock velocity $v_s = 5000$\kms\ is obtained from broad \ha\ and 
  He\,I 5876\,\AA\ \citep{Chugai_2019}.
Noteworthy, the CR acceleration efficiency reported in the latter paper is 
  overestimated by a factor of two.
In the case of SN~2002ic CS velocities $u = 80$\kms\ and $u_{max} = 250$\kms\ are 
 reported by \cite{Kotak_2004}, while the forward shock velocity $v_s = 6000$\kms\ 
 is obtained from the spectrum modelling \citep{CCL_2004}.
 
The efficiency of CR acceleration for three SNe~IIn inferred using Equation (\ref{eq:precur}) 
  is given in Table \ref{tab:precur}.
Surprisingly, all the values are close to each other within a factor of 1.5.
This can be considered as a validation of the proposed diagnostics of the CR precursor 
  of SNe~IIn.
On the other hand, the universality of the CR acceleration efficiency for three SNe~IIn 
  picked at random is not a trivial fact given a possible variation of environmental conditions.

\begin{table}
\centering
\caption[]{CR acceleration efficiency for SNe~IIn}
\begin{tabular}{p{1.0cm}|p{0.8cm}|p{0.8cm}|p{0.8cm}|p{0.8cm}|p{0.8cm}}
\hline
 SN     &  Day     &    $u^{\dag}$    &  $u_{max}$  &   $~~v_s$  &   $~~\epsilon$ \\
\hline
 2008iy  &  761    &  45     &  145    & 2600    & 0.038 \\
 2002ic  &  256    &  80     &  250    & 6000    & 0.028 \\
 1997eg  &  198    &  20     &  160    & 5000    & 0.028 \\
 \hline
 \parbox[]{6cm}{\small $^{\dag}$ Velocity in km\,s$^{-1}$.}
 \end{tabular}
\label{tab:precur}
\end{table} 

\subsubsection{SN~2008iy radio emission} 

The found CR acceleration efficiency can be applied to the interpretation of the 
   SN~2008iy.  
The VLA observations at 8.46 GHz (3.5 cm) on 2009 April 24 (639 d after explosion) did not detect
 radio flux, whereas on 2009 December 26 (885 d) the radio was detected 
   with the flux $f_{\nu} \approx 192\pm36$\,$\mu\mbox{Jy}$  \citep{Chandra_2009}.
This means that on day 639 the CSM was opaque to $ff$-absorption, while on day 885 CSM 
  optical depth to $ff$-absorption was less then unity. 
  
This conjecture can be verified using our model of the CSM and assuming that the $ff$-absorption   
  is related to the intercloud CSM with the temperature of $T_e = 10^6$\,K.
The latter assumption is based on the presence of narrow coronal [Fe\,X] 6374\,\AA\ line 
  in the latest spectrum on day 711 after discovery \citep{Miller_2010}. 
The situation in this regard is similar to SN~1997eg \citep{Chugai_2019}.
For SN~2008iy the calculated optical depth to $ff$-absorption at $\lambda = 3.5$\,cm 
   is $\tau_{ff} = 4.05$ on day 639 
  and $\tau_{ff} = 0.57$ on day 885, which explains the evolution of radio flux 
    between these two epochs.
At the distance $D_L = 179$\,Mpc \citep{Miller_2010} the detected flux corresponds to the 
  unabsorbed monochromatic luminosity  $L_{\nu} = 1.3\times10^{28}$\ergs\,Hz$^{-1}$.

The interaction model at this age suggests the CDS radius $r_{cds} = 3.5\times10^{16}$\,cm, 
  preshock density $\rho_0 = 5\times10^{-17}$\,g\,cm$^{-3}$, and the CDS velocity 
  $v_{cds} = 2500$\kms.
The energy density of  cosmic rays is 
  $U_c = 3\epsilon \rho_0v_s^2 \approx 0.36$\,erg\,cm$^{-3}$ with $\epsilon \approx 0.038$.
The energy density of relativistic electrons is 
  $U_e = 3.6\times10^{-3}(R_{e,p}/0.01)$\,erg\,cm$^{-3}$, 
  where $R_{e,p} \sim 10^{-2}$ is the typical fraction of the relativistic electron in the 
   cosmic ray pressure produced largely by relativistic protons. 
 For the power law distribution of relativistic electrons $dN/dE = KE^{-p}$ with $p = 2$ 
   the synchrotron emissivity \citep{Getmantsev_1952} for the isotropic distribution of relativistic electrons is
 \begin{equation}
 4\pi j = 4\times10^{-18}K\lambda^{1/2}B^{3/2}~\mbox{erg\,s}^{-1}\mbox{cm}^{-3}\,.
 \end{equation}
Adopting $E_1 = 1$\,MeV and  $E_2 = 10^4$\,MeV one gets   
$K = U_e/\ln{(E_2/E_1)} = 3.9\times10^{-4}(R_{e,p}/0.01)$\,erg\,cm$^{-3}$.
The volume of the radio-emittiing shell is 
 $V = (4\pi/3)r^3\omega = 9\times10^{49}(\omega/0.5)$\,cm$^3$ with 
  $\omega \sim 0.5$ \citep[e.g.][]{Chevalier_1998} that corresponds to the shell 
  thickness of $\Delta r = 0.14r_{cds}$.
The monochromatic radio luminosity at $\lambda = 3.5$\,cm is thus $L_{\nu} =4\pi jV =  
   2.5\times10^{29}B^{3/2}(\omega/0.5)(R_{e,p}/0.01)$\,\ergs\,Hz$^{-1}$ which 
   fits the observational luminosity for $B = 0.14(\omega/0.5)^{-2/3}(R_{e,p}/0.01)^{-2/3}$\,G. 
Interestingly, this value is only a factor of two lower comared to estimates for SN~1986J and  
   SN~1988Z inferred using effects of the synchrotron self-absorption at low frequences \citep{Chevalier_1998}.
  
Moreover, the value $B \sim0.14$\,G is close to the saturated turbulent magnetic field 
  predicted by the DSA theory.  
Indeed, the predicted energy density of the saturated turbulent magnetic field 
  \citep{Bell_2004} is  
 \begin{equation}
 U_m \sim 0.5\left(\frac{v_s}{c}\right)U_c\,,
 \label{eq:mag}
 \end{equation}
   where $v_s$ is the forward shock speed, $c$ is speed of light. 
For SN~2008iy on day 885 with $v_s \approx v_{cds} = 2500$\kms\ and 
   $U_c = 0.36$\,erg\,cm$^{-3}$   Equation (\ref{eq:mag}) implies $B_{sat} \sim 0.19$\,G. 
The magnetic field in the radio-emitting shell of SN~2008iy thus coincides within a 
factor of unity with the saturated field suggested by the DSA mechanism.   
Note, for the estimated magnetic field of SN~2008iy the optical depth of the radio-emitting 
   shell to the synchrotron self-absorption on day 885 at 8.46\,GHz is small, 
   $\tau_{ssa}\sim 0.14$, and therfore does nor affect our interpretation of radio 
   emission.
  
To summarize, the recovered efficiency of cosmic ray acceleration in SN~2008iy 
  combined with the CS interaction model 
  is consistent with the radio emission of SN~2008iy for reasonable value of the 
   magnetic field.

  
\section{Discussion and Conclusions}
\label{sec:disc}

The paper has been aimed at the interpretation of the optical 
   phenomena related to the CS interaction of SN~2008iy.
The bolometric light curve is found to be powered by the collision of the 
  high energy ($3\times10^{51}$\,erg) SN ejecta  
  with the $\sim 10$\msun\ CS envelope residing  at the radius of $\sim 3\times10^{16}$\,cm.
The maximal mass loss rate of presupernova is 
   $\sim 0.09$\msun\,yr$^{-1}$ that took place $\sim200$ yr before the explosion. 
These estimates are based on the wind velocity 45\kms\ inferred from the 
  modelling of the CS \ha\ in the Keck I spectrum on day 711 \citep{Miller_2010}.
  
The origin of enormous mass loss rate that closely precedes the explosion is puzzling.
It looks like that the preSN  envelope "knows" what is going on in the core either 
  via pulsations caused by the hydrodynamic perturbations generated in the pre-collapse 
  core \citep{Shiode_2014}   
  or, possibly, via a merger process that favours the vigorous mass loss and should 
  end up with the core collapse.   
  
While the progenitor nature is enigmatic, the large mass of CSM and enormous 
  mass loss rate favour the massive progenitor with the main 
  sequence mass probably exceeding $30$\msun. 
In this context the former conjecture on the origin of SN~1988Z progenitor 
  from stars in the range 8-10\msun\ \citep{Chugai_1994} requires revision.
This reassessment is also supported by 8.5\,yr-long observations of SN~1988Z in 
   the optical and X-ray band, which suggest enormous the radiation output 
    $\sim 10^{51}$\,erg (for $H_0 =70$\kms\,Mpc$^{-1}$) and the CS mass $\gg 1$\msun\ \citep{Aretxaga_1999}.
Noteworthy that the high explosion energy of SN~2008iy is on the verge of possibiities of 
 the neutrino-driven  mechanism that can provide $\lesssim 2\times10^{51}$\,erg \citep{Janka_2017};  
  the rotation energy of a collapsing core mediated by the magnetic field therefore 
  is probably involved in the energetic explosion.

Based on the spectral similarity between SN~2008iy and SN~1988Z emphasised by 
  \cite{Miller_2010} the luminous \ha\ line is attributed to the emission 
  from CS clouds shocked and fragmented in the forward shock. 
The ionization of shocked clouds can be related to three sources: 
   (i) X-rays from the forward shock,(ii) X-rays 
    from cloud shocks, and (iii) extreme  UV radiation from the boundary layer of  
    cold dense fragments heated by the electron conductivity flux from 
    the hot gas of the forwad shock.
Without detailed modelling of the complex physics we are unable so far 
   to rank these mechanisms according to their significance -- all of them 
    can contribute to \ha\ luminosity.
  
The \ha\ smootheness with the fluctuation amplitude $\lesssim 1$\% suggests that the number 
  of line-emitting  fragments of CS clouds in the forward shock should be $\gtrsim 10^6$.
Although presumed, the fragmentation process of shocked clouds in SNe~IIn has never 
  been probed observationally.
Here the fluorescent  O\,I 8446\,\AA\ emission line is used for the first time 
  to demonstarate along with other observational constrants that 
   the minimal length scale of line-emitting cloud fragments is  
    2.3 dex smaller compared to the minimal size of the undisturbed CS clouds. 
 This should be considered as a convincing confirmation of the 
   fragmentation cascade accompaniing the CS clouds disruption in the forward shock. 
 
 It should be admitted that the ejecta interaction with cloudy CSM is more complicated phenomenon 
  than the presented scenario.
 The issue that requires elaborate study is the structure of the postshock layer between the 
   forward shock and the CDS.
 The CDS Rayleigh-Taylor (RT) instability \citep{CheBlo_1995,BloEll_2001} creates an extended 
  boundary layer at the 
   CDS interface with the forward shock filled in by the mixed RT spikes of a dense CDS material. 
 In that case CS clouds or/and fragments may not be in time 
   to completely mix in the postshock flow of the forward shock but instead 
    interact with this dense boundary layer. 
At first glance this effect should not change significantly major results but for generality 
  this process should be included in the detailed interaction 
  scenario \citep[e.g.][]{Jun_1996}.
  
The analysis of the narrow \ha\ reveals the effect of preshock acceleration of CS gas 
  identified with the outcome of CR precursor.
The detection of this effect in CS \ha\ of three random sample of SNe~IIn 
  (SN~2008iy, SN~1998eg, SN~2002ic) is of  paramount significance by two reason.
First, we acquire the simple efficient diagnostic tool to probe the CR acceleration efficiency 
  in late time ($\gtrsim 200$\,d) SNe~IIn based on high resolution 
  ($\lesssim 50$\kms) \ha. 
Second, comparable values of the CR acceleration efficiency (0.038, 0.28, 0.28) indicate 
  weak sensitivity of $\epsilon$ value from environmental conditions in SNe~IIn.

The synchrotron radio emission of SNe~IIn,  and young supernovae, in general,  clearly
   demonstrates that the CS interaction is accompanied by the CR acceleration and magnetic field 
   amplification \citep{Chevalier_1982x,Chevalier_1998}.
Note, however, that the radio data do not permit us to infer the CR ray acceleraton efficiency 
   since the relativistic electrons is a minor constituent of the CR and their emissivity also depends on the unknown  magnetic field.
This emphasises the unique role of the diagnostics of the CR 
  precursor based on the analysis of the resolved profile of CS \ha\ narrow line. 

Noteworthy that the interaction model combined with the recovered efficiency of CR acceleration 
   is consistent with the evolution of the radio emission of SN~2008iy, while the 
   inferred magnetic field is compatible with the general prediction of the magnetic field 
   amplification in the DSA theory.

\section*{Acknowledgements}

I am grateful to Adam Miller for the Keck spectrum of SN~2008iy.
  
\section*{Data Availability}

Details of the data analysis are available on request  
  
\bibliographystyle{mnras}
\bibliography{SN2008iy_ref} 
\bsp	
\label{lastpage}
\end{document}